\documentclass[conference]{IEEEtran}
\usepackage{cite}
\ifCLASSINFOpdf
\else
\fi
\usepackage{amsmath}
\usepackage{algorithmic}
\usepackage{array}
\usepackage{fixltx2e}
\usepackage{stfloats}

\usepackage{enumitem}
\usepackage{amsfonts}
\usepackage{pifont}
\usepackage{algorithm}
\usepackage{graphicx}
\usepackage{graphics}
\usepackage{multirow}
\usepackage{balance}
\usepackage{hyperref}

\newcommand{\cmark}{\ding{51}}
\newcommand{\xmark}{\ding{55}}

\begin{document}
\title{InTAR: Inter-Task Auto-Reconfigurable Accelerator Design for High Data Volume Variation in DNNs}

% \author{

% \IEEEauthorblockN{Zifan He}
% \IEEEauthorblockA{University of California, Los Angeles \\
% zifanhe1202@g.ucla.edu}
% \and
% \IEEEauthorblockN{Anderson Truong}
% \IEEEauthorblockA{University of California, Los Angeles \\
% andersonleetruong@gmail.com}
% \and
% \IEEEauthorblockN{Yingqi Cao}
% \IEEEauthorblockA{University of California, San Diego \\ yic033@ucsd.edu}
% \and
% \IEEEauthorblockN{Jason Cong}
% \IEEEauthorblockA{University of California, Los Angeles \\ cong@cs.ucla.edu}
% }


\author{
\IEEEauthorblockN{
Zifan He\IEEEauthorrefmark{1},
Anderson Truong\IEEEauthorrefmark{1},
Yingqi Cao\IEEEauthorrefmark{2} and
Jason Cong\IEEEauthorrefmark{1}}

\IEEEauthorblockA{\IEEEauthorrefmark{1}University of California, Los Angeles}
\IEEEauthorblockA{\IEEEauthorrefmark{2}University of California, San Diego\\
Email: zifanhe1202@g.ucla.edu, andersonleetruong@gmail.com, yic033@ucsd.edu, cong@cs.ucla.edu}
}


\maketitle

\begin{abstract}
The rise of deep neural networks (DNNs) has driven an increased demand for computing power and memory. Modern DNNs exhibit high data volume variation (HDV) across tasks, which poses challenges for FPGA acceleration: conventional accelerators rely on fixed execution patterns (dataflow or sequential) that can lead to pipeline stalls or necessitate frequent off-chip memory accesses. To address these challenges, we introduce the Inter-Task Auto-Reconfigurable Accelerator (InTAR) \footnote{\url{https://github.com/OswaldHe/InTAR}}, a novel accelerator design methodology for HDV applications on FPGAs. InTAR combines the high computational efficiency of sequential execution with the reduced off-chip memory overhead of dataflow execution. It switches execution patterns automatically with a static schedule determined before circuit design based on resource constraints and problem sizes. Unlike previous reconfigurable accelerators, InTAR encodes reconfiguration schedules during circuit design, allowing model-specific optimizations that allocate only the necessary logic and interconnects. Thus, InTAR achieves a high clock frequency with fewer resources and low reconfiguration time. Furthermore, InTAR supports high-level tools such as HLS for fast design generation. We implement a set of multi-task HDV DNN kernels using InTAR. Compared with dataflow and sequential accelerators, InTAR exhibits $\mathbf{1.8\times}$ and $\mathbf{7.1 \times}$ speedups correspondingly. Moreover, we extend InTAR to GPT-2 medium as a more complex example, which is $\mathbf{3.65 \sim 39.14\times}$ faster and a $\mathbf{1.72 \sim 10.44\times}$ more DSP efficient than SoTA accelerators (Allo and DFX) on FPGAs. Additionally, this design demonstrates $\mathbf{1.66 \sim 7.17\times}$ better power efficiency than GPUs.
\end{abstract}


% \fi
\IEEEpeerreviewmaketitle

\section{Introduction} \label{sec:intro}

% \begin{figure}[ht]
%     \centering
%     \includegraphics[width=0.85\columnwidth]{imgs/intra-arch.png}
%     \caption{Example timeline of dataflow and sequential execution patterns and \textsc{InTAR}. Intermediate data volumes between tasks are indicated as proportional to the area of the white rectangles. Only sequential execution requires off-chip memory access.}
%     \label{fig:intra-arch}
% \end{figure}

The development of deep neural networks (DNNs) has driven numerous breakthroughs in AI-assisted applications \cite{dong2019unified, dosovitskiy2020image} while intensifying demands for compute efficiency due to increasing network complexity and memory requirements. As DNN applications grow in size \cite{touvron2023llama, raiaan2024review}, larger data volumes generated between operations strain hardware resources during deployment. Consequently, a key challenge in designing DNN accelerators is: \textit{Given resource constraints imposed by physical hardware limitations or power considerations, how can execution scheduling be improved to minimize end-to-end latency while accounting for hardware overhead?}

A common strategy of DNN accelerator designs is grouping computations belonging to the same matrix/vector operation into a \textbf{task} and scheduling computations at the task level. For example, self-attention \cite{vaswani2017attention} of transformer models involves two tasks: $QK^T$ and softmax operation. Previous accelerators focus on two types of fixed task execution patterns: \textbf{dataflow} (also called spatial) \cite{chen2024allo, basalama2023flexcnn} or \textbf{sequential} (also called temporal) executions \cite{jouppi2023tpu, bai2023fet, liu2021hardware, zhang2020dnnexplorer}. For \textbf{dataflow} executions, each processing element (PE) specializes in a single task. Data is streamed through FIFOs connecting PEs for dependent computations with minimal on-chip buffer for data reuse. Depending on the dataflow graph, it includes both task-pipeline and task-parallel executions. \textbf{Task-pipeline} streams data between dependent tasks, and \textbf{task-parallel} executes independent tasks in parallel. For \textbf{sequential} executions, a versatile PE processes tasks serially and parallelizes the computations inside each task. Resources are reused across operators. Although both execution patterns exhibit strong performance, each of them faces a major challenge for large DNN applications \cite{chen2024understanding, zhuang2024ssr}:
\begin{itemize}[wide=0pt]
    \item \textbf{Dataflow: reduced computation resource efficiency}. Due to the coarse and fine-grained pipeline stalls from the data dependencies, some PEs in dataflow execution may remain idle, leading to low computation resource efficiency and high end-to-end latency. 
    \item \textbf{Sequential: high off-chip memory access overhead.} Output data from previous tasks will be stored and reused by the next tasks in sequential execution. Moreover, data may be cached and remain unused until the consumer tasks start executions. Due to the limited on-chip resources, it is often difficult to store all output data in SRAM. If storing data in the off-chip memory, the accelerator will suffer from high latency and energy consumption of off-chip memory access.  
\end{itemize}

\begin{table}[ht]
  \centering
  \caption{Minimum and Maximum Intermediate Data Size of Sample DNNs}
  \label{table:trans-cnn}
  \resizebox{0.8\columnwidth}{!}{
  \begin{tabular}{|l|l|l|l|}
    \hline
    \textbf{Model} & \textbf{Type} & \textbf{Min Data Size} & \textbf{Max Data Size}\\
    \hline
    GPT 2 \cite{radford2019language} & Transformer & $64L$ & $\max(4096L, L^2)$ \\
    \hline
    Llama 2 7B \cite{touvron2023llama} & Transformer & $128L$ & $\max(22016L, L^2)$ \\
    \hline
    ResNet-50 \cite{he2016deep} & CNN & 151K & 802K \\
    \hline
    ResNet-152 & CNN & 151K & 802K \\
    \hline
    ResNext-101 \cite{xie2017aggregated} & CNN & 151K & 1.04M \\
    \hline
  \end{tabular}
  }
\end{table}

To address both challenges simultaneously, we observe another commonly neglected feature of DNNs to exploit: \textbf{high data-volume variation} (HDV). Table \ref{table:trans-cnn} lists a set of DNN applications and the minimum/maximum input/output data size between operations. $L$ is the sequence length for transformer models. These models offer both a large maximum data size (up to 88 MB with $L$ = 2048) and a significant data size variation ($5.3\times \sim 172\times$ from minimum to maximum data size). A potential solution to improve computation resource efficiency while circumventing off-chip memory access for intermediate data is to execute sequentially for tasks that produce small data within on-chip memory capacity, and stream data for tasks that produce large data.

Therefore, we propose a novel accelerator design paradigm on FPGAs: \textbf{inter-task auto-reconfigurable accelerator} (\textsc{InTAR}). \textsc{InTAR} can switch execution patterns automatically based on on-chip memory and computation resources. When a task produces large intermediate data, \textsc{InTAR} pipelines multiple tasks to avoid accessing off-chip memory for this data. Otherwise, \textsc{InTAR} will process tasks sequentially to maximize computational efficiency by eliminating pipeline stalls. Compared with other reconfigurable accelerators \cite{cai2023inter, cong2014fully, baek2020multi, liu2022overgen, kong2023hardware}, \textsc{InTAR} allows \textit{model-specific circuit optimization} that keeps only necessary control logics and interconnects for reconfiguration. Hence, InTAR requires fewer reconfiguration resources, achieves a high clock frequency, and has a low reconfiguration overhead (10 to 20 ns) (Section \ref{sec:intrra}). Moreover, since computations are reconfigured at the task level, \textsc{InTAR} is one of the first works regarding FPGA-based reconfigurable accelerators that support high-level hardware generation tools such as High-Level Synthesis (HLS) \cite{cong2022fpga} for fast accelerator design. Specifically, our contributions include:

\begin{itemize}[wide=0pt]
    \item We present \textsc{InTAR}, a novel accelerator design paradigm on FPGAs that balances the memory access and computational efficiency tradeoff for HDV applications (Section \ref{sec:intrra}). 

    \item We illustrate the techniques to design \textsc{InTAR} with HLS and important considerations in placement and routing for hardware implementation of \textsc{InTAR} on FPGAs (Section \ref{sec:design}).
    
    \item Evaluations. We implement \textsc{InTAR} for five multi-task kernels that broadly exist in DNN applications (Self-Attention, Multi-layer CNN, FFN layer, VAE, and Gating Network) to show its advantages. InTAR exhibits $\mathbf{1.8\times}$ and $\mathbf{7.1 \times}$ speedup compared with the corresponding dataflow and sequential accelerator. We further present InTAR on the GPT-2 medium model for a complete DNN example, which achieves a speedup of $\mathbf{3.65 \sim 39.14\times}$ and a $\mathbf{1.72 \sim 10.44\times}$ improvement in DSP efficiency compared to the SoTA accelerators (Allo \cite{chen2024allo} and DFX \cite{hong2022dfx}). Moreover, \textsc{InTAR} demonstrated $\mathbf{1.66 \sim 7.17\times}$ better power efficiency compared to GPUs.
\end{itemize}

% Resource constraints come from:
% 1. physical hardware constraints: you cannot use more than what the FPGA board supports.
% 2. Energy pre-design optimization: The designer may use tools like Xilinx power design manager to find out the resource configuration given a power constraint. Thus, we have to follow the new constraint when designing the accelerator.

\section{Background and Motivations}

\subsection{Dataflow and Sequential DNN Accelerators}

Dataflow accelerators allocate resources per DNN task and pipeline computations to conserve memory while preserving performance. For example, FlexCNN \cite{basalama2023flexcnn} composes CNN-based accelerators in HLS, GenGNN \cite{abi2022gengnn} accelerates GNNs, and Allo \cite{chen2024allo,chen2024understanding} designs FPGA accelerators by parsing Python scripts into HLS code for each operator, automatically combining layer optimizations with customization primitives. 

On the other hand, sequential accelerators process layers in DNN sequentially and attempt to utilize all available resources. GPUs and TPUs \cite{jouppi2023tpu} are considered instruction-based sequential accelerators that support various DNNs. ZyncNet \cite{gschwend2020zynqnet} is a CNN accelerator deployed on Zync SoC with a specialized topology. FlightLLM \cite{zeng2024flightllm} is a vector processor implemented on Alveo U280 FPGA for Llama model inference. 

\subsection{Hybrid Accelerators}

Some previous works attempt to mitigate the drawbacks of dataflow and sequential accelerators by having a hybrid execution pattern. A naive approach is allocating part of the resource for dataflow execution and the rest for sequential execution \cite{zhang2020dnnexplorer}. A more recent method is to allocate several PEs and schedule multiple tasks to the same PE. For instance, SSR \cite{zhuang2024ssr} is a hybrid accelerator implemented on AMD Xilinx Versal ACAP devices for vision transformers. By reserving two groups of AI engines for two types of matrix multiplies, SSR allows sequential executions of matrix multiplies (GEMM) to minimize latency and enables data forwarding between dependent GEMMs to reduce the memory cost. However, PE specializations of hybrid accelerators impede the flexibility of execution pattern switching, which may lead to stalls. Section \ref{sec:case} will further discuss the inefficiency of hybrid accelerators with an example.

\begin{figure*}[ht]
    \centering
    \includegraphics[width=0.95\linewidth]{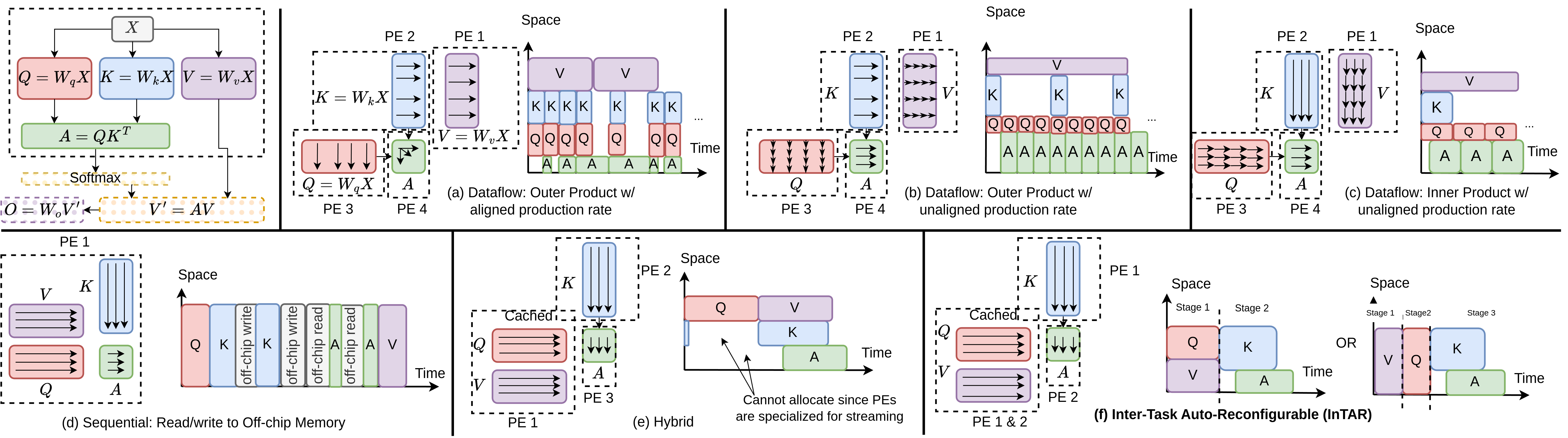}
    \caption{Example of mapping computations of attention and linear projection of value matrix to the dataflow, sequential, hybrid accelerators, and \textsc{InTAR}. The sequence length is 256, the hidden dimension is 1024, and the weights are $1024\times 1024$ matrices. Depending on the scheduling of the rest of the computations in the entire model, \textsc{InTAR} can choose between task-parallel (left) and sequential modes (right) for better locality. Both will have the same latency.}
    \label{fig:attn_sample}
\end{figure*}

\subsection{Reconfigurable Accelerators}

A reconfigurable accelerator can change its microarchitecture on the fly. A common implementation of such accelerators is the coarse-grain reconfigurable array (CGRA) \cite{liu2022overgen, cong2014fully, tanomoto2015cgra}, which compiles a dataflow graph into a configuration and modifies switches connecting PEs accordingly. However, CGRA has complex reconfiguration logic to ensure generalizability. This introduces substantial area overheads \cite{taras2019impact} and additional wiring that negatively affects the timing \cite{liu2022overgen}.

An alternative method to reconfigure resources is dynamic partial reconfiguration (DPR) \cite{vipin2018fpga}, which reserves regions on FPGAs to load partial bitstreams for runtime reconfiguration. While having a low resource overhead for the configuration controller, it has a higher reconfiguration overhead (10 to 100 ms) than FPGA-based CGRA (around 10 ns).

Other reconfigurable accelerators focus on optimization under only specific computation scenarios \cite{baek2020multi, cai2023inter, tong2024FEATHER}. For example, SET \cite{cai2023inter} performs inter-layer scheduling of convolution networks onto the tiled accelerators utilizing a time-space resource-allocation tree. FEATHER \cite{tong2024FEATHER} mainly resolves reduction efficiency in each CNN layer by reconfiguring the reduction dataflow.

Considering these limitations, existing reconfigurable accelerators can hardly apply to a broad range of modern DNNs, which necessitates the development of a new design paradigm. 

\subsection{Applications with High Data Volume Variation}

A multi-task application is defined as high data-volume variation (HDV) if the input/output data sizes between tasks are highly varied. Many of the DNNs are HDV for two reasons:
\begin{itemize}[wide=0pt]
    \item Modern DNNs have complex dependencies. Data is produced several layers in advance and can only be discarded after they are reused. Hence, the data production and elimination rates vary widely.
    \item The objectives of many DNN workloads involve feature extraction (e.g., classification) that reduces the data size, or data construction (e.g., image/text generation), which increases the data size. By the nature of these workloads, the data size will change during execution.
\end{itemize}

% Why do most data-volume-variant applications belong to the machine learning field?
% 1. modern machine learning applications have deep networks with complex dependencies. Data produced in previous layers may be reused in the following layers (e.g. skip connection, gating network). Data production/elimination rate can vary from layer to layer.
% 2. Nature of machine learning tasks: many ML tasks involve feature extraction (e.g. classification tasks) with down-sampling, data constructions (e.g., image/text generation) with up-sampling, or both (e.g., GAN, VAE, Transformers)

\subsection{A Motivating Case Study: Attention + Linear Projection} \label{sec:case}

We provide a case study to illustrate the advantages of inter-task reconfiguration for HDV DNNs: calculating the attention score and a linear projection of the value matrix. Figure \ref{fig:attn_sample} shows the dataflow graph of the attention layer and the timeline for each execution pattern. The direction of the arrows indicates the output production order, and the lengths represent the production rate. Here we focus on calculating $Q$, $K$, $V$, and $A$ in the dashed square. In this example, we assume that:
\begin{itemize}[wide=0pt]
    \item Matrix multiplies are weight-stationary as weights are already stored in on-chip memory. 
    \item Outputs ($A$ and $V$) will be written into off-chip memory.
    \item The rest of the on-chip memory size will only store either $Q$ or $K$, but not both. 
    \item Input sequence is short, which means $A$ can be stored in on-chip buffer with $K$ or $Q$.
\end{itemize}

The pattern of data size variation is \textit{first increase then decrease}. Dataflow execution instantiates PEs for each task. There are three ways to stream and cache data across tasks, illustrated in Figure \ref{fig:attn_sample}: outer product with aligned and unaligned production rates of $Q$ and $K$ (Figure \ref{fig:attn_sample}(a) and (b)), and inner product with unaligned production rates (Figure \ref{fig:attn_sample}(c)). Inner products with aligned production rates require caching of both $Q$ and $K$, which is infeasible. Each case incurs compute resource idling due to either varied workload size of $A$, back-pressure from streaming, or early completion of tasks.

Sequential execution instantiates a single PE to execute each task serially, as shown in Figure \ref{fig:attn_sample}(d). Due to insufficient on-chip memory capacity, it needs to read from and write to the off-chip memory, increasing the overall latency. 

Some hybrid accelerators \cite{zhuang2024ssr} can mitigate the issues by executing $Q$ and $V$ sequentially and pipeline $K$ and $A$. However, since PEs are specialized for either streaming data or cached buffer \cite{zhuang2024ssr}, computation of $V$ cannot be assigned to PE 2 and PE 3, and these two PEs are idle at the beginning, waiting for the arrival of data for $Q$.

For \textsc{InTAR} (Figure \ref{fig:attn_sample}(f)), PEs can be reconfigured to either sequential or dataflow execution at different times. Therefore, PEs can be efficiently allocated while off-chip memory access for intermediate data is avoided. Section \ref{sec:intrra} discusses how \textsc{InTAR} manages this in detail.

\section{Inter-Task Auto-Reconfigurable Accelerator} \label{sec:intrra}

To improve resource efficiency, \textsc{InTAR} minimizes off-chip memory access by employing dataflow execution when necessary while keeping the rest of the execution as sequential as possible. The finest granularity of execution pattern switching is the \textbf{task}, which represents a group of computations belonging to the same matrix or vector operation. Execution pattern switching is achieved through automatic reconfiguration based on instructions stored in the static configuration buffer (Section \ref{sec:templ}), which is hardened into the design. These instructions trigger computation behavior and data movement changes. 

Unlike other reconfigurable \cite{liu2022overgen, cong2014fully, tong2024FEATHER, cai2023inter, kong2023hardware} and overlay accelerators \cite{yu2019opu, yu2020light}, which prioritize abstraction and fast compilation, InTAR focuses on reconfiguration for optimization. The reconfiguration schedule in InTAR is determined at \textbf{circuit design time} rather than relying on post-design external commands. For every application, a \textbf{static schedule} is created based on the specific application and resource constraints, and the circuit is then tailored accordingly. This approach reduces reliance on general-purpose control units and cross-bars, retaining only essential reconfiguration logic and interconnects. The result is low resource utilization for reconfiguration controls (49\% LUT reduction for a design with $3\times$ throughput compared to \cite{liu2022overgen} for multiple GEMMs), high clock frequency ($2.41\times$ that of \cite{liu2022overgen}), and low reconfiguration latency ($10^6\times$ faster than DPR \cite{kong2023hardware}, and comparable to \cite{liu2022overgen}) compared with FPGA-based CGRA and DPR approaches. Additionally, InTAR handles reconfiguration at the task level, enabling rapid and convenient development with high-level tools like HLS. This makes InTAR well-suited for FPGAs to adapt quickly to new DNN applications. Lastly, InTAR's reconfiguration scope is not limited to specific computations. Table \ref{tab:reconfig_comp} compares InTAR with prior DNN accelerator works.

% In this section, we first illustrate \textsc{InTAR}'s execution modes and how \textsc{InTAR} achieves high resource efficiency in the case study by switching between modes. Then, we describe the architecture template of \textsc{InTAR}. 

% Difference between domain-specific CGRA/Overlay and InTAR and the benefits:
% 1. Interconnect and resource overhead when having fixed and generalized microarchitecture for various applications.
% 2. Focus on a single or a fixed set of applications with a single microarchitecture design. It cannot extend to future applications with more complicated dependencies and data volume variance.

% The convenience of implementing InTAR: since scheduling is handled at task level, high-level synthesis (HLS) is sufficient to implement a high-quality InTAR design, allowing hardware designers at the HLS level and compiler developer to leverage InTAR without touching RTL code (example HLS code after).

\subsection{Execution Modes of \textsc{InTAR}} \label{sec:comp-mode}

\begin{figure*}[ht]
    \centering
    \includegraphics[width=0.9\textwidth]{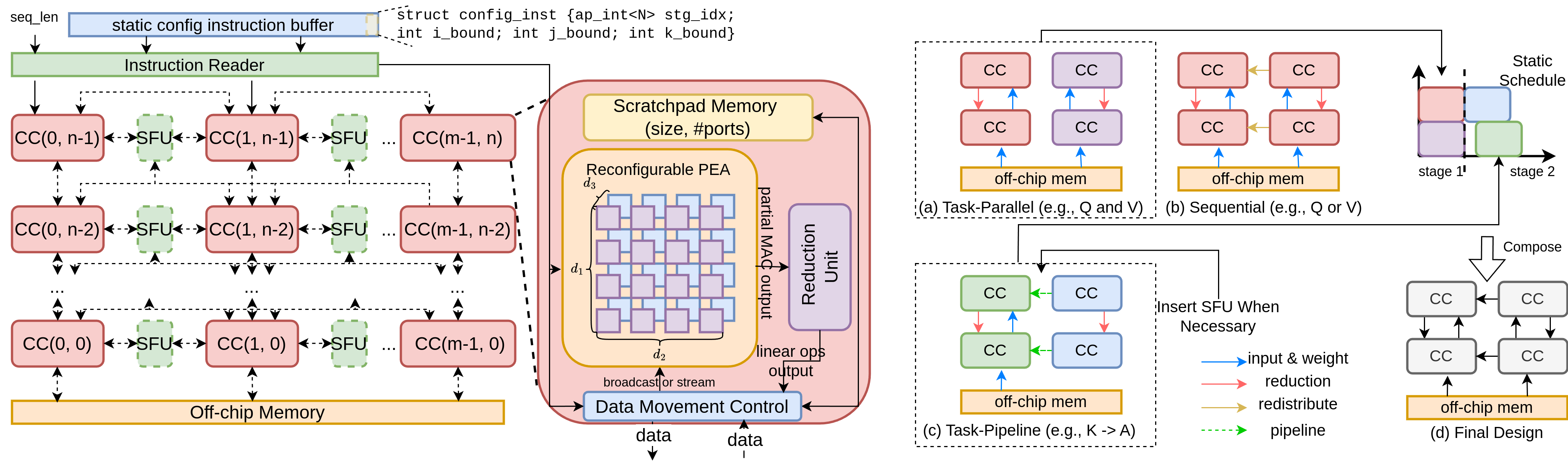}
    \caption{Left: Architecture template of \textsc{InTAR}. Compute cores (CC) compute linear operations (e.g., GEMM, ConvNet), and SFUs compute non-linear operations (e.g., softmax, GeLU). Each CC contains a scratchpad memory, a reconfigurable MAC unit array, a reduction unit, and a data movement control unit. Dashed lines indicate the candidates for connection between CCs and SFUs. Right: example architectures for each execution mode within the template, with $n,m=2$.  For a schedule, we will compose the architectures of the required modes to keep only the necessary interconnects and logic.}
    \label{fig:intrra-template}
\end{figure*}

\begin{table*}[ht]
    \centering
    \caption{Comparison between \textsc{InTAR} and other DNN acceleration works. The major reason for having idle compute resources is described in the parentheses for each work.}
    \resizebox{\textwidth}{!}{
    \begin{tabular}{|c|c|c|c|c|c|c|c|c|c|}
        \hline
        \textbf{Prior works} & \textbf{Platform} & \textbf{Contribution Type} & \textbf{Accelerator} & \textbf{Inter-task} & \textbf{Idle Compute} & \textbf{Intermediate} & \textbf{When Reconfig} & \textbf{Reconfiguration} &\textbf{Model-spec.} \\
        & & & \textbf{Category} & \textbf{Schedule} & \textbf{Resource Exist?} & \textbf{Data Movement} &  \textbf{Determined} & \textbf{Design Scope} &  \textbf{Design Opt.}\\
        \hline
        Allo \cite{chen2024allo} & FPGA & Domain-Spec. Lang. & Dataflow & \xmark & Yes (dataflow execution)& On-chip only & - & - & \xmark\\
        \hline
        FQ-BERT \cite{liu2021hardware}, DFX \cite{hong2022dfx} & FPGA & Accelerator & Sequential & \xmark & No & On-/Off-chip & - & - & \xmark \\
        \hline
        SSR \cite{zhuang2024ssr} & FPGA & Accelerator & Hybrid & \cmark & Yes (PE specialization) & On-chip only & - & - & \xmark\\
        \hline
        % Herald \cite{kwon2021heterogeneous} & ASIC & Hybrid & \cmark & \xmark & Yes (single DNN) & Off-chip & - & - & \xmark \\
        % \hline
        FPCA \cite{cong2014fully}, OverGen \cite{liu2022overgen} & FPGA & Architecture & Reconfigurable & \xmark & Yes (single DNN) & On-/Off-chip  & After circuit generation & General & \xmark \\
        \hline
        FEATHER \cite{tong2024FEATHER} & FPGA & Accelerator & Reconfigurable & \xmark & No & On-/Off-chip & After circuit generation & Reduction Network & \xmark \\
        \hline
        SET \cite{cai2023inter} & ASIC & Scheduler & Reconfigurable & \cmark & Yes (sub-opt. sched.) & On-/Off-chip & N/A & Tiled Accelerator & \xmark \\
        \hline
        \hline
        \textbf{\textsc{InTAR}} & FPGA & Design Paradigm & Reconfigurable & \cmark & No & On-chip only & Circuit design time & General & \cmark \\
        \hline
    \end{tabular}
    }
    \label{tab:reconfig_comp}
\end{table*}

Depending on the available resources on FPGAs and the properties of tasks, \textsc{InTAR} configures the resources into sequential, task-pipeline, or task-parallel mode, corresponding to the dataflow and sequential execution patterns mentioned in Section \ref{sec:intro}. 
\begin{itemize}[wide=0pt]
    \item \textbf{Sequential}: All resources perform a single task in parallel.
    \item \textbf{Task-pipeline}: Same as dataflow execution with pipelined dependent tasks. Computational resources are partitioned proportionally to the workloads. Data are streamed from one task to another using FIFOs. 
    \item \textbf{Task-parallel}: Same as dataflow execution with independent tasks running in parallel. The resource efficiency is the same as the sequential mode, but it may have a better locality for subsequent tasks (e.g., for the example in upper left of Figure \ref{fig:attn_sample}, if calculating $V'$ and $O$ are pipelined, then running $Q$ and $V$ in parallel and assigning $V$ to the PE that will compute $V'$ avoids data aggregation overhead from other PEs). 
\end{itemize}

Figure \ref{fig:attn_sample}(f) illustrates the task assignment and timeline of \textsc{InTAR} for the case study in Section \ref{sec:case}. PEs 1 and 2 handle both attention computation and linear projections. Depending on the subsequent operations, \textsc{InTAR} can choose either task-parallel or sequential mode for $Q$ and $V$. If the scheduler decides to serialize computing $V'$ and $O$, then the sequential mode is more suitable to allow PEs 1 and 2 to execute cooperatively. If the scheduler pipelines computations of $V'$ and $O$, then task-parallel mode is preferred to improve locality. Both choices will not affect the latency but will determine the data movement overhead of the following tasks. After calculating $Q$ and $V$, \textsc{InTAR} switches to task-pipeline mode to compute $K$ and $A$, since storing $Q$ consumes all scratchpad memory.

% \begin{figure*}[ht]
%     \centering
%     \includegraphics[width=0.8\linewidth]{intra_csra_flow.png}
%     \caption{An overview of the CDS flow. Circles represent the tasks in a dataflow graph, or operations in DNNs. The same color indicates the same operations.}
%     \label{fig:csra}
% \end{figure*}

\subsection{Architecture Template} \label{sec:templ}

Figure \ref{fig:intrra-template} is the architecture template of \textsc{InTAR} for HDV DNN applications, which consists of a grid of compute cores (CC) computing linear operations and special function units (SFU) inserted in between. Each CC contains:
\begin{itemize}[wide=0pt]
    \item Reconfigurable PE Array (PEA): a multi-dimensional array of PEs that allows changing inputs and precisions. Data communication between PEs is either in a systolic array style or by broadcasting the data to all PEs. Each PE produces the complete or partial result of the linear operation.
    \item Scratchpad Memory: store intermediate and weight data.
    \item Reduction Unit: read the partial outputs from the PEA if needed and perform accumulations. 
    \item Data Movement Control Unit: determine the flow of data to the scratchpad memory and other CCs/SFUs.
\end{itemize}
When switching execution modes, a global instruction reader will read both the configuration instructions from a static buffer and the input size passed as the kernel argument. It will then modify part of each instruction related to the input size (e.g., the loop bounds) and send it to the CCs. The reconfigurable PEA and data movement control will read the instructions and propagate them among the CCs. The instruction sequence is determined by the static schedule of mode switching at the circuit design time. Each instruction contains a \textbf{stage index} and several\textbf{ loop bounds} (e.g., reconfiguring the three loop bounds for GEMMs). A stage refers to the execution of a single or a group of tasks that follows one of the execution modes, seen in Figure \ref{fig:attn_sample}(f), e.g., computing $Q$ and $V$ in stage 1 and $K$ and $A$ in stage 2. Each CC has a specified behavior for each stage, orchestrated by multiplexers with stage indices as the select signals. Notice that the stage-specific behaviors vary from one application to another. Thus, \textsc{InTAR} only needs to handle the reconfigurations defined in that application-specific schedule. The reconfiguration overhead consists of four parts: reading, modifying, sending, and decoding instructions to generate signals for multiplexers in the CCs. Each of these is done in a single cycle. Thus, reconfigurations have a 4-cycle overhead (10 to 20 ns, depending on frequency), which is negligible compared to modern DNN inference latency. 

The template has several design parameters for users to derive a specific architecture, including:
\begin{itemize}[wide=0pt]
    \item Number of rows $n$ and columns $m$ of the CCs. 
    \item Instantiation of the SFUs, which determine the positions and computations of each SFU.
    \item FIFO connections between the CCs and SFUs.
    \item Scratchpad memory size and number of memory ports.
    \item Reconfigurable PE array dimensions.
\end{itemize}
Depending on the schedule, we assign values to these parameters to perform the three execution modes mentioned in Section \ref{sec:comp-mode}. The right of Figure \ref{fig:intrra-template} depicts the example architectures of each execution mode derived from the template for $n,m=2$ for the case study in Section \ref{sec:case}. The colors of connections indicate the purpose of communications (read input and weights, perform reductions, redistribute output, or pipeline intermediate results). For a specific schedule, we compose the derived architectures of the execution modes utilized to get the final circuit design. \textbf{Common interconnections and control logic across execution modes will be merged, and unused interconnects will never be introduced.}

% \subsection{Challenges of \textsc{InTAR}}

% We identify two challenges that will impede the advocating of \textsc{InTAR} as an accelerator design paradigm on FPGAs:

% \noindent \textbf{Fast and convenient design flow.} Previous work in reconfigurable accelerator on FPGAs \cite{liu2022overgen, cong2014fully, tong2024FEATHER} relies on time-consuming RTL flow, but supports multiple applications with a single bitstream. Since \textsc{InTAR} requires design per application, it is crucial to have a faster and more convenient design flow rather than adopting the RTL flow. This will make designing large applications more manageable.

% \noindent \textbf{Complex logic required to repurpose resource on FPGAs.} To generate a place-able and routable design on FPGA with high clock frequency, having utilization under resource constraints is far from enough. Especially for designs with auto-reconfiguration, the control logic is usually complex and difficult to achieve timing closure.

\section{Designing Accelerators in \textsc{InTAR}} \label{sec:design}

The first step in designing \textsc{InTAR} accelerators is performing design space exploration (DSE). Given a dataflow graph (DFG) and device constraints, one searches through various combinations of architecture parameters and task execution modes (see Section \ref{sec:intrra}) to minimize total latency.

\subsection{Finding Schedule and Architecture Template Parameters}
In this work, we first topologically sort the tasks based on dependencies in the DFG and apply a heuristic-based DSE:
\begin{itemize}[wide=0pt]
    \item If the total size of output and previously cached data is larger than the memory constraint, then select the task-pipeline mode.
    \item Otherwise, if task-parallel mode can improve locality for the subsequent tasks as illustrated in Section \ref{sec:comp-mode}, then pick task-parallel mode. If not, then choose the sequential mode.
    \item Non-linear operations are pipelined with previous tasks.
\end{itemize}
Then, we analyze the generated schedule and the hardware configurations and determine architecture template parameters with the following heuristics: The schedule infers the instantiation of the SFUs and FIFO connections. The column count of CCs equals the maximum number of tasks of all stages, and the row count is the number of dies on the FPGA necessary to improve floorplanning. We compute the PEA dimensions to maximize utilization with limited cross-die communications and calculate the memory ports needed. Finally, the scratchpad memory is the maximum memory size required over all stages. For applications with variable input size, we schedule based on the maximum possible size and employ batching \cite{looks2017deep} to prevent inefficiency of the design for small data size. Although the heuristics are not guaranteed optimal, we show that our schedule results improved performance consistently (see Sections \ref{sec:testbench}, \ref{sec:gpt2}). We leave the development of an optimized DSE engine as future work.

\subsection{Designing Reconfigurable Modules: A Case Study in HLS}
In this work, we require the designer to manually write the reconfiguration designs based on the static schedule. Following the architecture template, there are three types of reconfigurations: data movement, compute, and control. All of them can be implemented in HLS efficiently with \textbf{conditional dataflow}, which utilizes if-else statements to declare the behavior changes and exploit HLS's resource binding pass to merge interconnects and logic as much as possible.

\begin{figure}[ht]
    \centering
    \includegraphics[width=0.8\columnwidth]{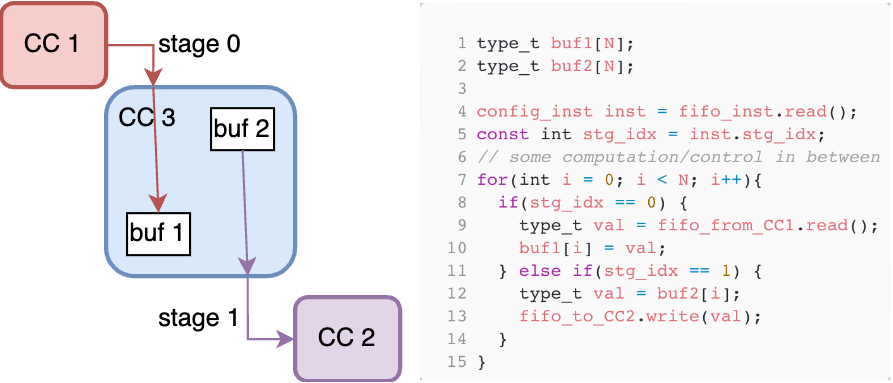}
    \caption{Example of data movement reconfiguration. Left: In different stages, PE 3 either read from PE 1 and write to on-chip buffer buf 1, or read from buf 2 and write to PE 2. Right: the corresponding HLS code.}
    \label{fig:hls_data_move}
\end{figure}

\noindent\textbf{Data Movement Reconfiguration}: a majority of auto-reconfigurations in \textsc{InTAR} are powered by data movement control. In different stages, each CC may load data from various sources (e.g., other CCs, on-chip buffers, registers, or constants) and send it to multiple destinations. Figure \ref{fig:hls_data_move} is an example of the data movement control reconfiguration in CC3: data flow from CC1 to buf 1 at stage 0 and flow from buf 2 to CC 2 at stage 1. The corresponding HLS code is a realization using the stage index as the condition to identify which flow the data movement control should adopt. Different stages can share data sources and destinations, which can potentially save interconnects. For instance, if the data flows from CC1 for each stage, the reader module can be reused.

\begin{figure}[ht]
    \centering
    \includegraphics[width=0.95\columnwidth]{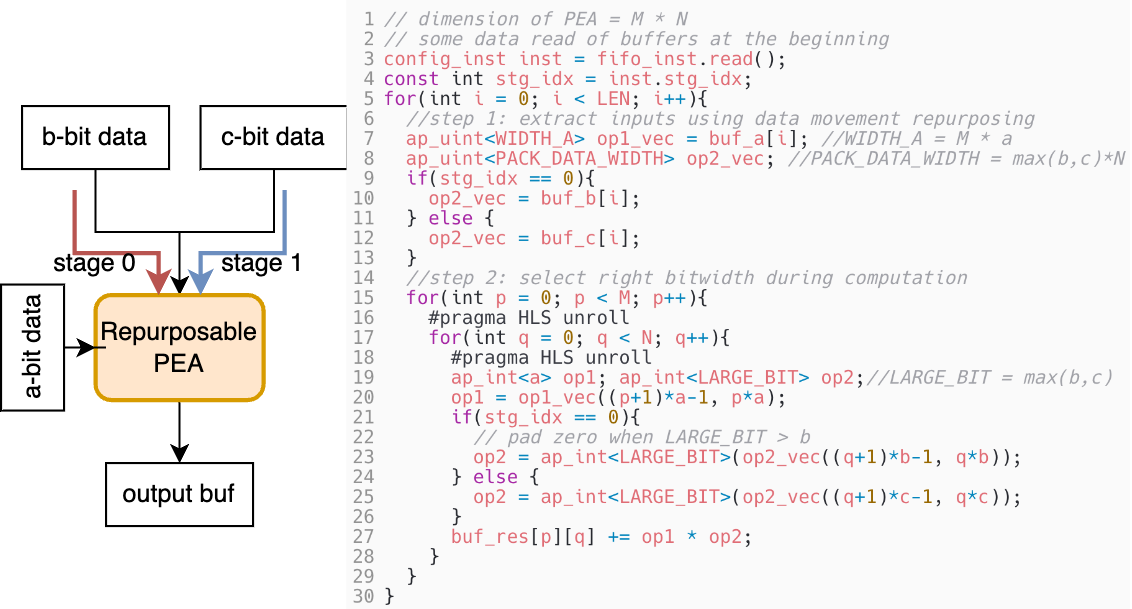}
    \caption{Example of compute reconfiguration. Left: The reconfigurable PEA consumes different buffers at each stage and supports mixed precision of b-bit and c-bit calculations. Right: the corresponding HLS code.}
    \label{fig:hls_compute}
\end{figure}

\noindent \textbf{Compute Reconfiguration}: Based on the definition of the architecture template, the reconfigurable PEA can change the input operands and the precision. Figure \ref{fig:hls_compute} illustrates a PEA switching the data source between stages, where each source is interpreted as a different precision. To change the input operands, we adopt a similar method of data movement reconfiguration to extract inputs from two source buffers conditioned on the stage index. Inputs are packed for data parallelism. Then, when computations start, each PE will select part of the input operand and pad with zeros to make sure the bitwidths are uniform across stages. This guides the HLS to bind compute resources between stages. The two steps cannot be merged, otherwise, the HLS will instantiate two PEAs with different precision specifications.

% // dimension of PEA = M * N
% // some data read of buffers at the beginning
% config_inst inst = fifo_inst.read();
% const int stg_idx = inst.stg_idx;
% for(int i = 0; i < LEN; i++){
%   //step 1: extract inputs using data movement repurposing
%   ap_uint<WIDTH_A> op1_vec = buf_a[i]; //WIDTH_A = M * a
%   ap_uint<PACK_DATA_WIDTH> op2_vec; //PACK_DATA_WIDTH = max(b,c)*N
%   if(stg_idx == 0){
%     op2_vec = buf_b[i];
%   } else {
% 	op2_vec = buf_c[i];
%   }
%   //step 2: select right bitwidth during computation
%   for(int p = 0; p < M; p++){
%     #pragma HLS unroll
%     for(int q = 0; q < N; q++){
%       #pragma HLS unroll
%       ap_int<a> op1; ap_int<LARGE_BIT> op2;//LARGE_BIT = max(b,c)
%       op1 = op1_vec((p+1)*a-1, p*a);
%       if(stg_idx == 0){
%         // pad zero when LARGE_BIT > b
%         op2 = ap_int<LARGE_BIT>(op2_vec((q+1)*b-1, q*b));
%       } else {
%         op2 = ap_int<LARGE_BIT>(op2_vec((q+1)*c-1, q*c));
%       }
%       buf_res[p][q] += op1 * op2;
%     }
%   }
% }

\begin{figure}[ht]
    \centering
    \includegraphics[width=0.9\columnwidth]{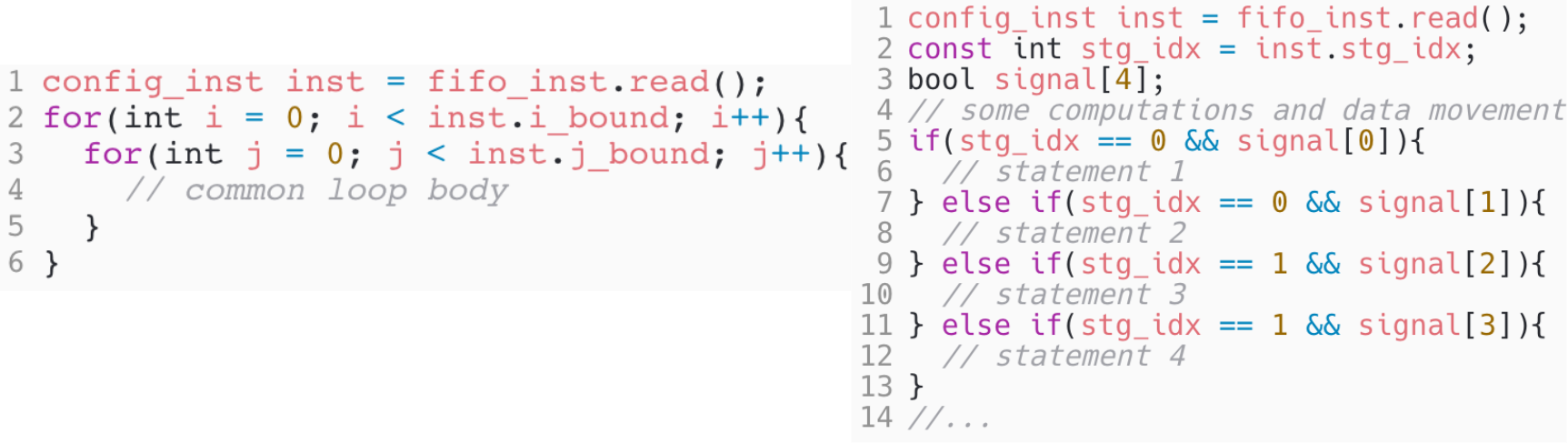}
    \caption{Example of control reconfigurations in HLS. Left: loop-bound control. Right: data-dependent conditional dataflow.}
    \label{fig:hls_control}
\end{figure}

\noindent \textbf{Control Reconfiguration}: In \textsc{InTAR}, there are control reconfigurations for loop bounds and data-dependent conditional dataflow. For loop-bound control, we can implement variable loop bounds for different stages of computations using the configuration instructions (left of Figure \ref{fig:hls_control}). For data-dependent conditional dataflow, when the behavior of executions depends on both stage index and other data sources, we can compose the conditions with logical operators (right of Figure \ref{fig:hls_control}). 

All accelerators in the \textsc{InTAR} paradigm are covered by the compositions of the three types of reconfigurations mentioned above, indicating complete support in HLS.

\subsection{Important Considerations in Placement and Routing}

In order to achieve rapid development, we need to make sure that the generated design is both valid and performant. Following are several important considerations of implementing a placeable and routable \textsc{InTAR} design with high frequency.

\noindent\textbf{Distribute Independent Tasks Across Dies.} For multi-die FPGAs \cite{Xilinx, versal}, cross-die wires are limited and often negatively affect the timing. Thus, when encountering a design with resource utilization higher than the single-die capacity, cross-die wiring latency can be the bottleneck of low clock frequency. A simple heuristic is to place independent tasks (e.g., multi-head attentions) into different dies instead of executing with all resources to avoid cross-die communications. Since there is no data movement between these tasks, we also do not add FIFOs between these tasks, which further reduces the cross-die wiring. 

\begin{figure}[ht]
    \centering
    \includegraphics[width=0.6\columnwidth]{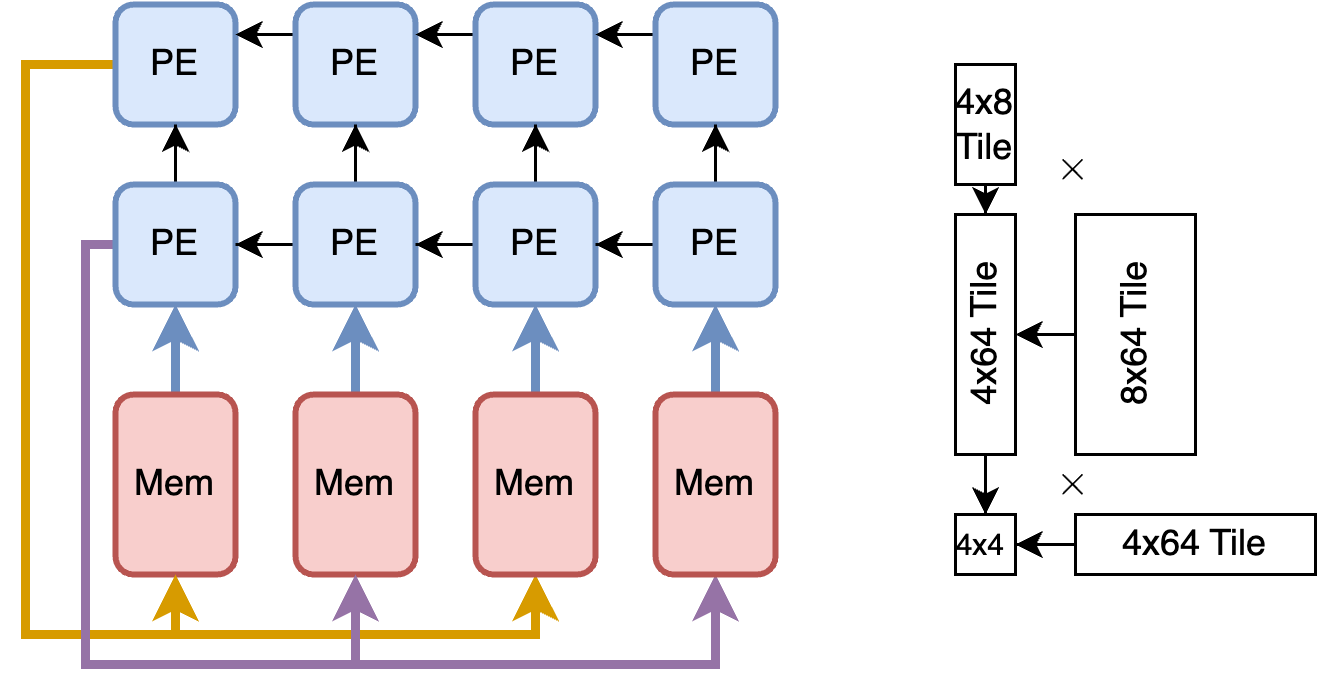}
    \caption{Example of a design with high-fanout nets on write ports of on-chip memory banks due to unaligned parallel memory access dimensions.}
    \label{fig:mem_part}
\end{figure}

\noindent\textbf{Matching On-chip Memory Parallel Access Dimension.} To serve multiple MAC units in each PEA, scratchpad memory is partitioned by access patterns. In \textsc{InTAR}, the PEA in the CCs may require varying patterns between tasks. For example, Figure \ref{fig:mem_part} shows two dependent GEMMs: the first computes 2048 MACs simultaneously ($4 \times 8 \times 64$), and the second computes 1024 MACs ($4 \times 4 \times 64$). Since the PEA is reused across tasks, we aim to equalize the throughputs of the two matrix multiplies for efficiency, necessitating further partitioning of the second GEMM. This causes high-fanout nets at the write ports, increases cycle time, and reduces frequency. We address this by tiling operands so that output tiles have equal dimensions (e.g., $16\times16$), aligning parallel read/write dimensions even after matrix transpose. 

\begin{table*}[ht]
    \centering
    \caption{Descriptions of multi-task kernels with the corresponding parameters and data size constraints. The on-chip memory constraints are artificial to make the kernel HDV.}
    \resizebox{0.95\textwidth}{!}{
    \begin{tabular}{|p{0.09\textwidth}|>{\raggedright}p{0.23\textwidth}|>{\raggedright}p{0.18\textwidth}|>{\raggedright}p{0.17\textwidth}|p{0.2\textwidth}|p{0.17\textwidth}|}
        \hline
        \textbf{Kernel} & \textbf{Description} & \textbf{DNN Applications} & \textbf{Parameters} & \textbf{Data Size Constraints} & \textbf{Data Variation Pattern} \\
        \hline
        Self Attention (Attn) & A set of dependent matrix multiplications that extract the contextual relationships between tokens in a sequence. & Widely used in sequence modeling (Transformer \cite{vaswani2017attention}, ViT \cite{han2022survey}, GAT \cite{velivckovic2017graph}) & 256 token sequences. The hidden dimension is 1024. & Min data size: 0.5 MB, Max data size: 1.0 MB, On-chip memory constraint: 0.625 MB & Increase ($Q, K$) $\rightarrow$ Decrease ($Q, K \rightarrow A$) $\rightarrow$ Increase ($V$)$\rightarrow$ Decrease ($A, V \rightarrow V'$) \\
        \hline
        FFN Layer (FFN) & Three layers of linear projections for feature extractions. Input is a single vector. & Elementary block in many CV and LLM applications (ResNet \cite{he2016deep}, YoloV3 \cite{farhadi2018yolov3}, Transformer) & Input size = final output size = 256. First layer output size = 1024, second layer output size = 2048. & Min data size: 0.5 KB, Max data size: 2 KB, On-chip memory constraint: 1 KB & Increase $\rightarrow$ Increase $\rightarrow$ Decrease \\
        \hline
        Multi-layer CNN (M-CNN) & Four layers of convolutions with upsampling and pooling layers & Used in many CV applications (ResNet, VGG \cite{vedaldi2016vgg}) & Input size = 224, kernel size = 3, $2\times2$ upsampling and pooling size. & Min data size: 98 KB, Max data size: 1.57 MB, On-chip memory constraint: 125 KB. & Increase $\rightarrow$ Increase $\rightarrow$ Decrease $\rightarrow$ Decrease\\
        \hline
        Variational Autoencoder (VAE) & Contains an encoder in convolutions for the latent space distribution and a decoder in transpose convolutions. & Used for image compression and the generator model in GAN (VAE-GAN \cite{gao2020zero}) & Input size = 28, kernel sizes are 4 and 8, 2 channels and 2 filters & Min data size: 1.3 KB, Max data size: 3.1 KB, On-chip memory constraint: 1.5 KB &  Decrease (encoder 1) $\rightarrow$ Decrease (encoder 2) $\rightarrow$ Increase (decoder 1)$\rightarrow$ Increase (decoder 2) \\
        \hline
        Gating Network (GN) & Two parallel linear projections with element-wise product and a sequential linear projection. & Llama-family LLMs \cite{touvron2023llama} & Sequence length = 32, input dimension = 4096, hidden dimension = 11008 & Min data size: 0.25 MB, Max data size: 0.67 MB, On-chip memory constraint: 0.5 MB & Increase (up projections) $\rightarrow$ Decrease (element-wise product, down projection)\\
        \hline
    \end{tabular}
    }
    \label{tab:eval_benchmark}
\end{table*}

To evaluate \textsc{InTAR}, we will show its broadness in DNN accelerations (Section \ref{sec:testbench}) and its advantages when deploying more complex DNNs on different FPGAs (Section \ref{sec:gpt2}).

\section{Evaluation 1: Multi-Task Kernels in HDV DNNs} \label{sec:testbench}

\subsection{Experiment Setup}
To demonstrate the broadness of \textsc{InTAR} in HDV DNN accelerations, we construct a testbench with five multi-task kernels that can serve as a standalone DNN application (Multi-layer CNN, Variational Autoencoder) or participate in part of the DNN applications (Self Attention, FFN Layer, Gating Network). Table \ref{tab:eval_benchmark} illustrates these kernels in the evaluation. Unlike existing benchmarks \cite{zhou-rosetta-fpga2018, pouchet2012polybench} for FPGA-based accelerator evaluation, our testbench has three features crucial for HDV: \ding{172} it covers various DNN applications in computer vision, language modeling, graph processing, etc., displayed in the "DNN Applications" column; \ding{173} it adopts hyperparameters used in the real applications (e.g., Gating Network has the identical dimensionalities as Llama 2\cite{touvron2023llama}); and \ding{174} kernels are all HDV, and we impose a memory constraint between the minimum and maximum input/output data sizes so that \textsc{InTAR} can exploit its ability to switch execution modes. 

Using our testbench, we compare \textsc{InTAR} with human-optimized dataflow and sequential accelerators, each designed in four weeks. All designs are evaluated under identical resource constraints, including on-chip memory capacity, the number of DSPs, and off-chip memory bandwidth utilization. All designs are implemented utilizing Xilinx Vitis HLS 2021.2 with TAPA \cite{guo2023tapa} and evaluated on Alveo U280 FPGA board, with a clock frequency of 300 MHz. We chose the human-optimized designs as the baseline since: \ding{172} There is a lack of existing designs of the kernels in our testbench with the same model and hardware configurations, and \ding{173} existing frameworks \cite{Duarte:2018ite, basalama2023flexcnn, chen2024allo} for DNN application design generation on FPGAs do not handle off-chip memory communications or do not consider the extra resource constraints introduced to evaluate HDV DNNs. 

\begin{figure}[ht]
    \centering
    \includegraphics[width=0.95\columnwidth]{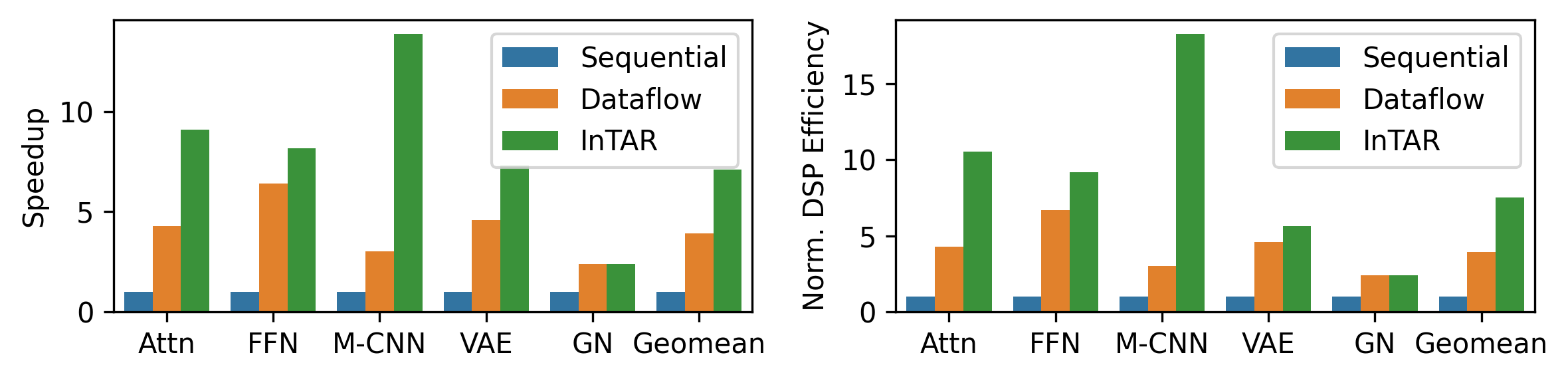}
    \caption{Speedup and DSP efficiency of \textsc{InTAR} over dataflow and sequential execution for the five multi-task kernels in the testbench. Values are normalized from sequential accelerators.}
    \label{fig:intrra-eval1}
\end{figure}

\subsection{Analysis}

Figure \ref{fig:intrra-eval1} shows the speedup and DSP efficiency of InTAR over sequential and dataflow accelerators. Overall, InTAR achieves a speedup of $7.1 \times$ and $1.8\times$ over sequential and dataflow accelerators, along with $7.5 \times$ and $1.9\times$ higher DSP efficiency (GOP/s/DSP) in geometric mean. The performance boost of InTAR over sequential accelerators mainly comes from the reduction of off-chip memory access, where InTAR has 76\% lower off-chip memory access volume than the sequential accelerators in geomean over the five kernels. For dataflow accelerators, InTAR is significantly faster in self-attention and multi-layer CNN than the other three kernels. There are two explanations. First, the two kernels mentioned have a higher data reuse between dependent tasks for matrix multiplications and convolutions, which requires data buffering and preprocessing before continuing the computations (e.g., the pooling layer of the multi-layer CNN will wait until producing two rows of outputs from the previous layer to downsample the data). Since the selected kernels have a long chain of dependent tasks, pipeline stall latency may dominate the end-to-end latency. Second, InTAR can better reuse off-chip memory ports for these two kernels. For example, after loading inputs and finishing the first convolution layer in the multi-layer CNN, InTAR can utilize the memory ports allocated to the inputs to write outputs. Thus, InTAR has a higher effective off-chip read/write throughput than dataflow accelerators, which only use the dedicated memory port for input/output. For Gating Networks, InTAR cannot further reduce latency compared with dataflow accelerators since only input and output data sizes are within the memory constraint, and our schedule for InTAR is identical to pipelining all tasks. 

\section{Evaluation 2: GPT-2 Input Prefilling} \label{sec:gpt2}

% \begin{figure}[ht]
%     \centering
%     \includegraphics[width=0.8\columnwidth]{imgs/intra_fpga_design_v2.png}
%     \caption{A sample design of InTAR mapped to Versal VPK180 FPGA for GPT-2 Medium model. (a) Overall architecture and parameters. Each SLR contains two CCs and one SFU. (b) The scheduled timeline. The color of the blocks corresponds to the PEs or SFUs shown in (a).}
%     \label{fig:u280-design}
% \end{figure}

\subsection{Experiment Setup}

We choose the GPT-2 medium input prefilling stage as a complex scenario to further evaluate \textsc{InTAR}, which is a frequently used benchmark by prior SoTA FPGA-based LLM accelerators \cite{chen2024allo, hong2022dfx}. We selected two platforms (Versal VPK180 \cite{versal-premium} and Alveo U280 \cite{Xilinx}), which result in different schedules and architecture template parameters based on their resource constraints. For instance, the design for VPK180 has four rows of CCs spreading out to four SLRs of the FPGA, while the design on U280 has 3 rows of CCs to fit its 3-SLR layout. Each SLR includes two CCs and an SFU for both platforms. Thanks to the grid structure of \textsc{InTAR}'s architecture template, we can adapt the design conveniently to other FPGA boards with only slight modifications. 

To generate the circuit design, we employ Xilinx HLS with the TAPA framework \cite{guo2023tapa} and prototype using the Vitis 2021.2 that performs well on older platforms such as U280, and Vitis 2024.1 optimized for VPK180. The floorplanning of the CCs and SFUs is predetermined based on the architectural template, while the remaining modules (e.g., auxiliary buffers) are managed by AutoBridge \cite{guo2021autobridge}. The designs are both placed and routed for U280 and VPK180. For comparison, we select Allo \cite{chen2024allo} and DFX \cite{hong2022dfx} as the SoTA accelerators, both of which utilize U280 as the compute platform; accordingly, we include InTAR implemented on U280.

% We also consider several other accelerators, but they are difficult to compare fairly. SSR \cite{zhuang2024ssr} targets the Versal AI engine and cannot deploy large transformer models due to limited on-chip memory. FEATHER's \cite{tong2024FEATHER} FPGA design does not contain the softmax module for the attention layer. FlightLLM \cite{zeng2024flightllm} only works for Llama models and has aggressive algorithmic optimizations, which are not comparable.

\begin{figure*}[ht]
    \centering
    \includegraphics[width=0.95\linewidth]{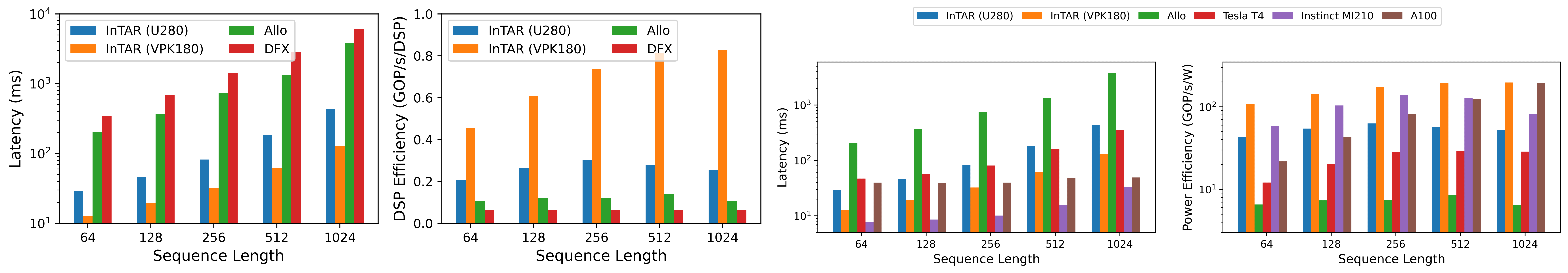}
    \caption{Left: Latency and DSP efficiency of \textsc{InTAR} (U280, VPK180), Allo, and DFX. Both designs of \textsc{InTAR} are significantly more DSP efficient. Right: Latency and power efficiency of \textsc{InTAR} (U280, VPK180), Allo \cite{chen2024allo}, and GPU solutions for GPT-2 medium model. }
    \label{fig:compare-gpu}
\end{figure*}

\begin{table}[ht]
    \centering
    \caption{Resource utilization and frequency of FPGA-based solutions (Allo, DFX, and \textsc{InTAR}) for GPT-2 medium input prefilling task. Speedup is normalized based on DFX.}
    \resizebox{0.85\columnwidth}{!}{
    \begin{tabular}{|c|c|c|c|c|}
        \hline
         & Allo \cite{chen2024allo} & DFX \cite{hong2022dfx} & \textsc{InTAR} & \textsc{InTAR} \\
        \hline
        \textbf{Device} & U280 & U280 & U280 & VPK180 \\
        \hline
        \textbf{Frequency} & 247 MHz & 200 MHz & 224 MHz & 300 MHz \\
        \hline
        \textbf{BRAM} & 389 (19\%) & 1192 (59\%) &535 (27\%) & 700 (14\%) \\
        \hline
        \textbf{DSP} & 1780 (20\%) & 3533 (39.2\%) & 6727 (75\%) & 6726 (47\%) \\
        \hline
        \textbf{LUT} & 569K (44\%) & 520K (40\%) & 485K (37\%) & 1074K (32\%) \\
        \hline
        \textbf{FF} & 653K (25\%) & 959K (43\%) & 627K (25\%) & 1072K (16\%) \\
        \hline
        \textbf{URAM} & 111 (12\%) & 104 (11\%) & 336 (35\%) & 412 (16\%) \\
        \hline
        \textbf{Norm. Speedup} & $1.83 \times$ & $1.0 \times$ & $14.64 \times$ & $39.14 \times$ \\
        \hline
    \end{tabular}
    }
    \label{tab:accel}
\end{table}

Table \ref{tab:accel} lists the resource utilization and frequency. \textsc{InTAR} on VPK180 has lower DSP utilization than U280 since VPK180 has fewer cross-SLR wires than U280 and PEA sizes are highly correlated with cross-SLR communications. DFX is executed in FP16, and all other designs employ the W4A8 format. We scale the DSP efficiency for DFX to align the data type. For performance metrics, we use OpenCL profiling functions to get the latency and \verb|xbutil| to measure the power within 100 runs on U280. Due to a lack of access to a physical device, we employ QEMU and the Xilinx Power Design Manager to calculate the latency and estimate the power consumption of VPK180. DFX is not included in comparing power efficiency due to a lack of data in the original work.

We also compare InTAR with various GPUs shown in Table \ref{tab:device}. PyTorch profiler measures the latency and memory transactions. To measure the power consumption, we employ the NVIDIA management library to probe the power every 10 ms and calculate the average power when it is stable. All GPUs are inference in BFloat16 format, which is a commonly supported and optimized data format for GPUs.

\begin{table}[ht]
    \centering
    \caption{Hardware configurations of the FPGAs and GPUs}
    \resizebox{\columnwidth}{!}{
    \begin{tabular}{|c|c|c|c|c|c|}
        \hline
         & U280 & VPK180 & T4 & A100 & MI210 \\
        \hline
        \textbf{Frequency} & 224 MHz & 240 MHz & 585 MHz & 765 MHz & 1.7 GHz \\
        \hline
        \textbf{Bandwidth} & 460 GB/s & 52.1 GB/s & 320 GB/s & 1.56 TB/s & 1.64 TB/s \\
        \hline
        \textbf{Peak Power} & 75 W & 180 W & 70 W & 250 W & 300 W \\
        \hline
        \textbf{Peak Perf.} & 8.09 TOP/s & 20.7 TOP/s & 65.13 TOP/s & 311.84 TOP/s & 181 TOP/s \\
        \hline
        \textbf{Process Node} & TSMC 16nm & TSMC 7nm & TSMC 12nm & TSMC 7nm & TSMC 6nm \\
        \hline
    \end{tabular}
    }
    \label{tab:device}
\end{table}

\subsection{Analysis} \label{sec:eval}

\textit{Comparison with FPGA-based Accelerators.} Figure \ref{fig:compare-gpu} presents the latency, DSP efficiency, and power efficiency comparison between FPGA and GPU accelerators. Compared to Allo and DFX, \textsc{InTAR} on U280 is $7.99 \times$ and $14.64 \times$ faster, and $2.19 \times$ and $3.98 \times$ more DSP efficient, respectively. \textsc{InTAR} on VPK180 is $21.39 \times$ and $39.14\times$ faster, and $5.66 \times$ and $10.44 \times$ more DSP efficient than Allo and DFX. \textsc{InTAR} on VPK180 is more DSP efficient than U280 since DSP58 has a higher throughput than DSP48. Additionally, implementing DSPs in dot-product modules on VPK180 trims the logic and further reduces cycles, leading to more than $2\times$ boost in DSP efficiency. Furthermore, similar to Evaluation 1 in Section \ref{sec:testbench}, we compared InTAR with our manually implemented dataflow and sequential accelerators, achieving $1.8\times$ and $8.3\times$ speedup and $1.7\times$ and $8.2\times$ DSP efficiency boost.

% \begin{table}[ht]
%     \centering
%     \caption{Performance comparison between \textsc{InTAR}, Allo, and FQ-BERT in throughput and DSP efficiency. The result for FQ-BERT is estimated for GPT-2 medium.}
%     \resizebox{\columnwidth}{!}{
%     \begin{tabular}{|c|c|c|c|}
%         \hline
%          & Allo \cite{chen2024allo} & FQ-BERT \cite{liu2021hardware} & \textsc{InTAR} (U280/VPK180) \\
%         \hline
%         \textbf{Model} & GPT-2 Medium & BERT-base & GPT-2 Medium \\
%         \hline
%         \textbf{Sequence Length} & 128 & 128 & 128 \\
%         \hline
%         \textbf{Latency (ms)} & 316.22 & 23.79 & 45.91 / 23.03 \\
%         \hline
%         \textbf{Throughput (GOP/s)} & 249.60 ($1.0\times$) & 939.47 ($3.8\times$) & 1719.23 ($6.9\times$) / 3427.26 ($13.7\times$) \\
%         \hline
%         \textbf{GOP/s/DSP} & 0.14 ($1.0\times$) & 0.29 ($2.1\times$) & 0.26 ($1.9\times$) / 0.51 ($3.6\times$) \\
%         \hline
%         \hline
%         \textbf{Model} & GPT-2 Medium & GPT-2 Medium$^*$ & GPT-2 Medium \\
%         \hline
%         \textbf{Sequence Length} & 512 & 512 & 512 \\
%         \hline
%         \textbf{Latency (ms)} & 1096.48 & 1152.65 & 184.01 / 73.65 \\
%         \hline
%         \textbf{Throughput (GOP/s)} & 305.57 ($1.0\times$) & 290.68 ($0.95 \times$) & 1820.88 ($6.0\times$) / 4549.36 ($14.9\times$) \\
%         \hline
%         \textbf{GOP/s/DSP} & 0.17 ($1.0\times$) & 0.09 $(0.53\times)$ & 0.28 ($1.6\times$) / 0.68 ($4.0\times$)\\
%         \hline
%     \end{tabular}
%     }
%     \label{tab:throughput}
% \end{table}

\textit{Comparison with GPUs.} As shown in Figure \ref{fig:compare-gpu}, \textsc{InTAR} on U280 achieves a $1.07\times$ speedup in geometric mean over T4 and $2.41\times$ speedup on VPK180. Especially for short sequences ($L \le 256$), InTAR’s performance surpasses both T4 and A100 GPUs. Moreover, with the same process node, \textsc{InTAR} on FPGAs attains higher power efficiency than GPUs. \textsc{InTAR} on U280 is $2.37 \times$ more efficient than NVIDIA T4. \textsc{InTAR} on VPK180 is $1.66\times \sim 7.17\times$ more efficient than T4, A100, MI210. One source of latency and power overhead for GPUs is the off-chip memory access. Since GPUs execute tasks sequentially and GPT-2 medium tends to generate large intermediate data, off-chip memory access is more intense on GPUs than \textsc{InTAR}-optimized FPGA designs. As a result, \textsc{InTAR} has a $20\% \sim 67\%$ lower off-chip memory access compared to T4, A100, and MI210.

\section{Conclusion and Future Work}

In this work, we propose InTAR, a novel accelerator design paradigm for DNNs with high data volume variation. InTAR enhances resource efficiency by switching execution patterns and performing model-specific optimizations. It surpasses SoTA FPGA accelerators in speed, DSP, and power efficiency, and, compared with GPUS, it reduces off-chip memory access.

Although focused on DNNs, InTAR may also enhance non-DNN HDV applications, which we plan to explore. Future work includes developing an optimized DSE engine to replace heuristic scheduling and automate InTAR design compilation.

\section*{Acknowledgement}

This work was supported in part by PRISM, one of the seven centers in the JUMP 2.0 program sponsored by SRC and DARPA. It is also supported by CDSC industrial partners and the AMD \footnote{J. Cong has a financial interest in AMD} HACC Program.



%\bibliographystyle{IEEEtran}

\bibliography{references}
\bibliographystyle{IEEEtran}

\end{document}